\documentclass[12pt]{article}
\usepackage{graphicx}
\usepackage[squaren,Gray]{SIunits}
\usepackage{booktabs}

\begin{document}
  \title{Optimized Si/SiO$_2$ high contrast grating mirror design for mid-infrared wavelength range:\\ robustness enhancement}
%
%
%
%

\author{C. Chevallier, F. Genty,  J. Jacquet \\
\textit{\footnotesize{Sup\'elec, 2 rue Edouard Belin 57070 Metz, France}}\\ 
\textit{\footnotesize{and}}\\ 
\textit{\footnotesize{LMOPS, Laboratoire Mat\'eriaux Optiques, Photonique et Syst\`emes}} \\
\textit{\footnotesize{ Unit\'e de Recherche Commune \`a l'Universit\'e Paul Verlaine - Metz et Sup\'elec}} \\
 N. Fressengeas \\
\textit{\footnotesize{LMOPS, Laboratoire Mat\'eriaux Optiques, Photonique et Syst\`emes}} \\
\textit{\footnotesize{ Unit\'e de Recherche Commune \`a l'Universit\'e Paul Verlaine - Metz et Sup\'elec}} \\
}


\maketitle
\begin{abstract}
A high reflectivity and polarization selective high contrast grating mirror has been designed with the use of an automated optimization algorithm. Through a precise study of the tolerance of the different lengths of the structure, the robustness with respect to the fabrication errors has been enhanced to high tolerance values between 5~\% and 210~\%. This adjustment of the dimensions of the structure leads to a 250~nm large bandwidth mirror well adapted for a VCSEL application at~$\lambda =$~\unit{2.65}{\micro\meter} and can easily be scaled for other wavelengths.
\end{abstract}
 \section{Introduction}

The Tunable Diode Laser Absorption Spectroscopy (TDLAS) is a sensitive and fast method for gas sensing used for instance to measure concentrations of polluting gas such as CH$_4$ and CO.
This technique requires the use of a tunable, stable and single mode operating source emitting in the mid infrared wavelength range beyond \unit{2}{\micro\meter} where strong absorption lines are present for these gas~\cite{chen_apb_2010, boehm_jcg_2010}. VCSELs are highly suitable for this kind of application, however, the development of continuous wave devices operating at room temperature at this wavelength range is still an important challenge.

Due to the small thickness of the gain region, VCSEL structures require high quality cavities with  highly reflective Bragg mirrors (R~$>$~99 \%). In the mid-infrared wavelength range, even though the best performances were obtained with AlGaInAsSb material system, the relatively low index contrast ($\Delta n \sim$~0.5) between DBR layers make the mirrors become as thick as \unit{11}{\micro\meter} impairing the electro-thermal-optical properties of the structure~\cite{cerutti_jcg_2009}. Currently, laser emission has been shown from an all-epitaxial monolithic microcavity near \unit{2.3}{\micro\meter} in a continuous wave mode and in quasi-CW (5~\%, \unit{1}{\micro\second}) up to \unit{2.63}{\micro\meter}~\cite{ducanchez_el_2009_2}. Another hybrid structure made of a dielectric top mirror and a buried tunnel junction operates in CW emission at room temperature in the 2.4-\unit{2.6}{\micro\meter} wavelength range~\cite{bachmann_njp_2009}.

One solution to increase the emission wavelength is the use of a high contrast grating mirror as top cavity mirror~\cite{huang_np_2007}. Such gratings, combined with a low index sub-layer, can exhibit high reflectivity of more than 99.9~\% for bandwidths larger than 100~nm~\cite{mateus_ptl_2004}. Moreover, due to their one dimensional symmetry, a high polarization selectivity can be performed by these mirrors and with a total thickness of less than \unit{2}{\micro\meter}~\cite{chevallier_apa_2010}, the stability and quality of the VCSEL emitted beam should be improved.

However, contrary to Bragg mirrors, the explanation of the optical response of high contrast grating (HCG) is not straightforward and the design adjustment made in order to achieve the required properties for a VCSEL application is still complex. Even though several formalism have been defined, based for instance on the destructive interference of modes~\cite{karagodsky_oe_2010}, the use of an optimization algorithm for the design combined with a numerical computation of the reflectivity keeps the advantage of versatility. Indeed, such a method easily allows the user to aim for specific properties of the HCG. It has been used for instance to design wide band and high diffraction efficiency grating~\cite{wang_ol_2010}, large bandwidth grating mirror for both TE and TM polarization~\cite{shokooh-saremi_oe_2008} or, like in this work, polarization selective large bandwidth mirror with technological constraints on the design dimensions~\cite{chevallier_apa_2010}.

In this work, we present a Si/SiO$_2$ HCG design optimized for a VCSEL application at $\lambda = $~\unit{2.65}{\micro\meter}. Then, through a precise study of the computed tolerances of the structure dimensions, the robustness is enhanced with respect to the fabrication errors.

 \section{Design of the HCG structure}

The high contrast grating structure studied in this work is based on Si/SiO$_2$ materials which properties and fabrication process are well known. The grating is made of silicium (n~=~3.435) on top of a low index layer of SiO$_2$ (n~=~1.509) used to achieved high reflectivity~\cite{mateus_ptl_2004} and allowing the use of a selective etching method for the fabrication of the grating. In order to increase the total reflectivity of the reflector and broaden the stopband, two quarter wavelength layers of Si/SiO$_2$ are combined with the grating~\cite{chevallier_apa_2010}. For the simulation of the design, the substrate is chosen as the VCSEL cavity material with an optical index of n~=~3.521.

In order to be well suitable for a VCSEL application, HCGs have to exhibit optical properties as we have defined with a 99.9~\% transverse magnetic (TM) reflectivity for the largest possible bandwidth. Moreover, to ensure a polarization stability of the emitted beam, the reflectivity of the transverse electric (TE) mode has been chosen to be kept lower than 90 \% for the whole bandwidth.

A 99.5~\% TM reflectivity should be enough to achieve laser emission but a 0.4~\% security margin  has been chosen to account for possible experimental growth imperfections and losses due to absorption. Indeed, the presence of OH radicals as impurities in SiO$_2$ results in an absorption band in the 2.6-\unit{2.9}{\micro\meter} range~\cite{soref_joa_2006}. In this reference, the absorption of silica has been measured with a 10 dB/cm value at \unit{2.65}{\micro\meter}, i.e. the refractive index becomes $n_{SiO2} = 1.509 + 2.1e^{-4} j$, which results in a 0.1~\% fall of the HCG reflectivity while a 0.4~\% decrease of reflectivity has been observed for a simulation with a 20 dB/cm attenuation coefficient  ( $n_{SiO2} = 1.509 + 1.7e^{-3} j$ ) at \unit{2.675}{\micro\meter}. This latter value appears therefore as the maximum absorption value allowed for the studied structure. However, the OH impurities concentration should strongly depends on the fabrication process and the exact value of the absorption must be determined in each case. For this theoretical study, a pure SiO$_2$ without any OH absorption was considered. Nevertheless, if the silica absorption is proven to be too high, this material can be replaced by another dielectric such as Si$_3$N$_4$.

The evaluation of the reflectivity of the mirror is made thanks to a rigorous coupled wave analysis (RCWA)~\cite{mrcwa, moharam_josaa_1995} which numerically finds an exact solution of Maxwell's equations for the electromagnetic diffraction problem of an infinite grating structure.

Reflection spectra for TE and TM modes are thus computed and through a well defined quality factor~\cite{chevallier_apa_2010}, the optical performance of the reflector is numerically evaluated. This performance is then maximized by a genetic optimization algorithm which searches for a global maximum of the quality factor by adjusting the design. The use of a global algorithm~\cite{openopt} is mandatory in this problem since the quality factor function exhibits numerous local maxima.

\begin{figure}
    \center\includegraphics[width=4.5cm]{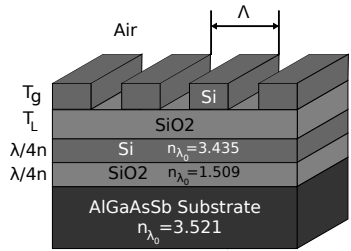}
    \caption{Scheme of the structure. The silicium grating defined by the empty length of the grooves $L_e$, the filled length $L_f$ and the grating thickness $T_g$ is on top of a SiO$_2$ layer of thickness $T_L$. These four parameters are adjusted to meet the characteristics of a VCSEL reflector.}
    \label{scheme}
  \end{figure}

In this work, the performances of the grating are optimized by adjusting the empty and filled lengths ($L_e$ and $L_f$), the grating thickness $T_g$ and the sublayer thickness $T_L$ as shown on Figure \ref{scheme}.

The use of an automated optimization presents the advantage of easily imposing technological constraints on the dimensions of the structure. These constraints have been set on the lengths $L_e$ and $L_f$ limited to a minimum value of 500~nm which should ease the photolithographic process. Besides, the etching process is also a critical step of the grating fabrication and, to achieve the theoretical grating profile presented on Figure~\ref{scheme}, the shape factor $SF = L_e / T_g$ of the grooves is kept at a minimum value of 0.9 since squared patterns are easier to etch than deep ones.

  \begin{figure*}
    \center\includegraphics[width=10cm]{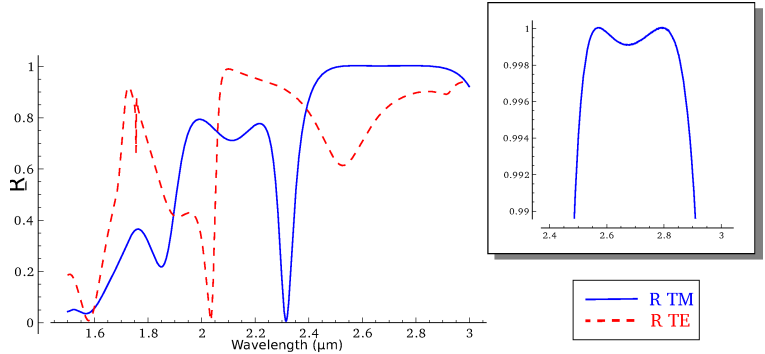}
    \caption{Reflection spectra for the TM mode (blue) and TE mode (dashed red) of the structure automatically optimized by a genetic-based algorithm. This optimum design exhibits a 307~nm large bandwidth with a 99.9~\% high TM reflectivity.}
    \label{optimum}
  \end{figure*}

The optimization for $\lambda = $~\unit{2.65}{\micro\meter} of the structure shown on Figure \ref{scheme} results in a 307~nm large bandwidth mirror exhibiting a high polarization selectivity with $R_{TE} < $~90 \% (Fig.~\ref{optimum}). The optimum lengths of the design are given in Table~\ref{table1} and satisfy the technological constraints with $L_e$~=~799~nm, $L_f$~=~541~nm and a shape factor $SF$~=~0.9. The grating thickness $T_g$~=~886~nm and the sublayer thickness $T_L$~=~321~nm combined with the two quarter wavelength layers result in a \unit{1.84}{\micro\meter} thick reflector.
Even though the performances of this HCG are well adapted for a VCSEL application, the tolerances of the dimensions of this grating are critical. For instance, the 884~nm minimum grating thickness leads to a maximum error allowed as small as 2~nm.

 \section{Optimization of the robustness}

\begin{table*}
  \caption{Tolerances of the optimum design found by the genetic-based algorithm.}
  \label{table1}
  \begin{center}
    \setlength{\tabcolsep}{0.5cm}
    \begin{tabular}{llll}

      \toprule\noalign{\smallskip}
      & Optimum & Minimum & Maximum\\
      \noalign{\smallskip}\hline\noalign{\smallskip}
      
      $L_e$ & 799 nm & 689 nm & 803 nm \\
      $L_f$ & 541 nm & 495 nm & 629 nm \\
      $T_g$ & 886 nm & 884 nm & 932 nm \\
      $T_L$ & 321 nm & 196 nm & 414 nm \\[0.2cm]

      $\Lambda = L_e + L_f$ & 1340 nm & 1260 nm & 1345 nm \\
      $FF = L_f / \Lambda$ & 40.37 \% & 35.97 \% & 45.67 \%\\[0.2cm]
      
      $\alpha_e = L_e + T_g$ & 1685 nm & 1652 nm & 1689 nm \\
      $\beta_e = L_e / \alpha_e$ & 52.58 \% & 50.80 \% & 54.89 \%\\[0.2cm]
      
      $\alpha_f = L_f + T_g$ & 1427 nm & 1388 nm & 1430 nm \\
      $\beta_f = L_f / \alpha_f$ & 62.09 \% & 59.13 \% & 64.74 \%\\
      \noalign{\smallskip}\bottomrule
    \end{tabular}
  \end{center}
\end{table*}
The evaluation of the tolerances is performed by computing the variation range of one parameter for which the HCG keeps a 99.9~\% TM reflectivity together with a $R_{TE} <$~90~\% at $\lambda_0 =$~\unit{2.65}{\micro\meter}. This computation is made by varying one parameter at a time, for instance the grating thickness $T_g$, while keeping the other ones ($T_L$, $L_e$ and $L_f$) constant at their optimal value.

The computation of the tolerances of the design found by the optimization algorithm exhibits large variation ranges of $\Delta L_e =$~14~\%, $\Delta L_f =$~25~\%, $\Delta T_g =$~5~\% and $\Delta T_L =$~68~\%. However, the optimum point is not centred in these variation ranges and the real tolerance values are much more limited with for instance a maximum increase of 4~nm on the empty length $L_e$. Moreover, these variation ranges are computed separately and give no information of the simultaneous error allowed on several parameters.

A solution to access this information is to compute the variation range of combinations of two dimensions. In the following, only ($L_e$,~$L_f$), ($L_e$,~$T_g$) and ($L_f$,~$T_g$) combinations are evaluated since the sublayer thickness $T_L$ is the most tolerant and centred parameter. Besides, $T_L$ does not depend on the etching process and should be fabricated with a better accuracy than the grating parameters.

  \begin{figure}
    \center\includegraphics[width=8cm]{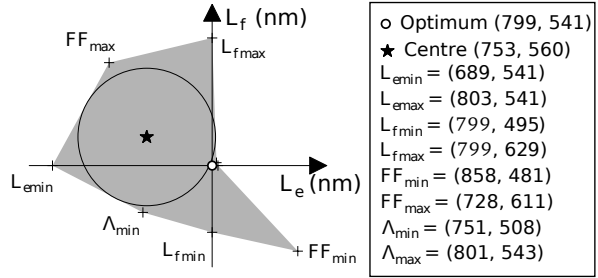}
    \caption{Tolerance map of $L_e$ and $L_f$. The variation ranges of the grating period $\Lambda = L_e + L_f$ and the fill factor $FF = L_f/\Lambda$ define a polygon (grey) of allowed ($L_e,~L_f$) couples for the design. The centre of the incircle ($\star$) enhances the robustness of the optimum point ($\circ$).}
    \label{LeLf}
  \end{figure}

For the first couple ($L_e$,~$L_f$), the variation ranges of two significant combinations of $L_e$ and $L_f$ are evaluated. The first one is the grating period given by $\Lambda = L_e + L_f$. This parameter is varied around its optimum value of $\Lambda = $~1340~nm while keeping the fill factor defined as $FF = L_e / \Lambda$ at its optimum value of 40.37~\%. The second combination of $L_e$ and $L_f$ is the fill factor $FF$ which variation range is evaluated with a constant grating period of $\Lambda = $1340~nm. Large tolerance values are exhibited by these parameters with 85~nm for $\Delta \Lambda$ and 9.7~\% for $\Delta FF$. The minimum and maximum of the variation ranges of the four parameters $L_e$, $L_f$, $\Lambda$ and $FF$, summarized in Table \ref{table1}, creating eight ($L_e$,~$L_f$) couples which can be plotted in a $L_f$ versus $L_e$ graph (Fig.~\ref{LeLf}). In the ($L_e$,~$L_f$) plan, these points define a polygon, in grey on Figure~\ref{LeLf}, which area represents the tolerance area of ($L_e$,~$L_f$)~couples allowed for the design.

The representation of the tolerances in a ($L_e$,~$L_f$) plan also shows very well the position of the optimum point within the tolerance area. In this case, the optimum point ($\circ$ on Figure~\ref{LeLf}) is located at an edge of the tolerance area ($L_e$~=~799~nm, $L_f$~=~541~nm). In order to increase the robustness regarding to the errors of fabrication which could be made on $L_e$ and $L_f$, the design should be centred within the tolerance area. To do so, the centre of the largest incircle of the polygon representing the tolerance area is computed. This point is thus the farthest point from any edge of the polygon. The centre ($\star$) on Figure~\ref{LeLf} corresponds to $L_e =$~753~nm and $L_f =$~560~nm and with a incircle radius of 47~nm, ensures a minimum tolerance of 94~nm on any combination of ($L_e$,~$L_f$).

  \begin{figure}
    \center\includegraphics[width=8cm]{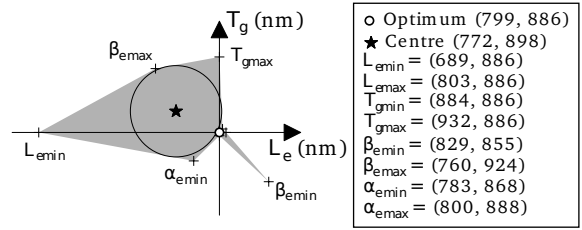}
    \caption{Tolerance map of $L_e$ and $T_g$. The variation ranges of $\alpha_e = L_e + T_g$ and $\beta_e = L_e/\alpha_e$ define a polygon (grey) of allowed ($L_e,~T_g$) couples for the design. The centre of the incircle ($\star$) enhances the robustness of the optimum point ($\circ$).}
    \label{LeTg}
  \end{figure}

The second dimension couple studied is the empty length and grating thickness ($L_e$,~$T_g$). In this case, both values of the optimum point (799,~886) are closed to the tolerance limit ($L_{e~max}$~=~803~nm, $T_{g~min}$~=~884~nm). This can easily be seen on Figure~\ref{LeTg} where the optimum is localized at the edge of the tolerance area. In this map, the first combination of ($L_e$,~$T_g$) is defined by $\alpha_e = L_e + T_g$ and the second one by $\beta_e = L_e/\alpha_e$. The polygon defined by the height extrema of the variation ranges of $L_e$, $T_g$, $\alpha_e$ and $\beta_e$ (Table~\ref{table1}) exhibits an incircle with a radius of 28~nm centred at (772, 898) making a tolerance of a minimum value of 56~nm on the lengths ($L_e$,~$T_g$) and any kind of their combinations.

\begin{figure}
    \center\includegraphics[width=8cm]{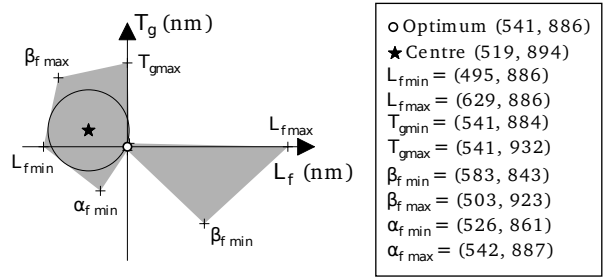}
    \caption{Tolerance map of $L_f$ and $T_g$. The variation ranges of $\alpha_f = L_f + T_g$ and $\beta_f = L_f/\alpha_f$ define a polygon (grey) of allowed ($L_f,~T_g$) couples for the design. The centre of the incircle ($\star$) enhances the robustness of the optimum point ($\circ$).}
    \label{LfTg}
  \end{figure}

The last couple, formed by the filled length $L_f$ and grating thickness $T_g$, is optimized by computing the tolerance values of $\alpha_f = L_f + T_g$ and $\beta_f = L_f / \alpha_f$. As shown on Figure~\ref{LfTg}, the polygon exhibits two large areas joined by a very thin path where the optimum point is located at (541,~886). Once again, the optimum point is in a critical location closed to several edges of the tolerance area. By choosing the centre of the incircle of the polygon, located at (519,~894), variation ranges of the lengths $L_f$ and $T_g$ can be increased to a 44~nm value.

 \section{Characteristics of the robust HCG}
    
\begin{table}
  \caption{Tolerances of the resulting design with optimized variation ranges.}
  \label{table2}
  \begin{center}\setlength{\tabcolsep}{0.3cm}
    \begin{tabular}{lllll}
      \toprule\noalign{\smallskip}
      & Optimum & Minimum & Maximum & \\
      \noalign{\smallskip}\hline\noalign{\smallskip}
      $L_e$ & 773 nm & 703 nm & 821 nm & (15 \%)\\
      $L_f$ & 541 nm & 504 nm & 608 nm & (19 \%)\\
      $T_g$ & 894 nm & 870 nm & 924 nm & (6 \%)\\
      $T_L$ & 321 nm & 0 nm & 674 nm & (210 \%)\\
      \noalign{\smallskip}\bottomrule
    \end{tabular}
  \end{center}
\end{table}

  \begin{figure*}
    \center\includegraphics[width=10cm]{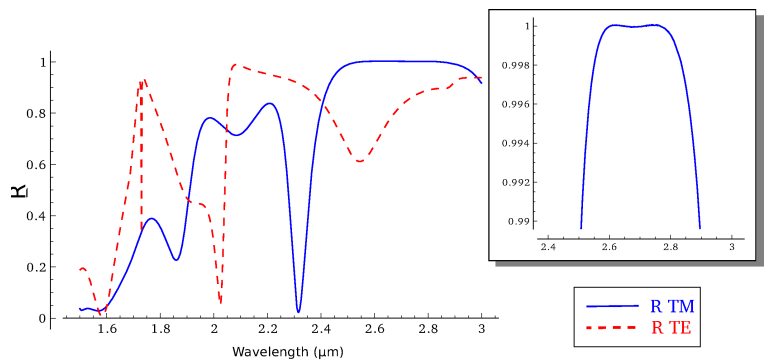}
    \caption{Reflection spectra for the TM mode (blue) and TE mode (dashed red) of the robust design with optimized tolerance values. This mirror exhibits a 250~nm large bandwidth for $R_{TM} >$~99.9~\% and between 5~\% and 210~\% of tolerance on its dimensions.}
    \label{centre}
  \end{figure*}
As a result of this optimization of the tolerances, three different designs are obtained. Each one is optimized to have the best tolerances for the couples ($L_e$,~$L_f$), ($L_e$,~$T_g$) and ($L_f$,~$T_g$). These three points represent a triangle in a three dimensional space ($L_e$,~$L_f$,~$T_g$). Thus, to find a unique design with optimized tolerances, the centre of the incircle of this triangle has been computed and result in a final optimized design with lengths $L_e = $~773~nm, $L_f = $~541~nm and $T_g = $~894~nm. This structure exhibits a 250~nm large bandwidth as shown on Figure~\ref{centre}, which is 59~nm less than the optimum one found by the genetic-based algorithm. However, the computation of the variation ranges of the design dimensions (Table~\ref{table2}) exhibits a very robust design with $\Delta L_e = $~15\%, $\Delta L_f = $~19\%, $\Delta T_g = $~6\% and $\Delta T_L = $~210\%. Each optimum dimension of the structure is well centred within the variation ranges which would ease the fabrication process.

 \section{Conclusion}
In this work, a Si/SiO$_2$ high contrast grating mirror has been designed for a VCSEL application at~\unit{2.65}{\micro\meter}. This mirror exhibits a 250~nm large bandwidth for $R_{TM} >$~99.9~\% together with a strong polarization selectivity by keeping $R_{TE} <$~90~\%. Moreover, with a total thickness of less than \unit{2}{\micro\meter}, such reflector should improve the quality of the emitted laser beam of the VCSEL.

From a technological point of view, the fabrication constraints are respected with a large pattern resolution ($>$~500~nm) and a grating shape factor closed to 0.9. Moreover, the robustness with respect to the fabrication errors has been enhanced and leads to tolerance values larger than 5~\% on the structure dimensions. Such characteristics make this HCG well adapted for a VCSEL application and should limit the pitfalls during the manufacturing process. Moreover, as spotted in~\cite{mateus_ptl_2004}, HCG can be scaled with wavelength in the limit of the refractive index dispersion.

 \section*{Acknowledgements}
The authors thank the French ANR for financial support in the framework of Marsupilami project (ANR-09-BLAN-0166-03) and IES and LAAS (France), partners of LMOPS/Sup\'elec in this project. This work was also  partly funded by the InterCell grant (http://intercell.metz.supelec.fr) by INRIA and R\'egion Lorraine (CPER2007).

\bibliographystyle{elsarticle/elsarticle-num}

\begin{thebibliography}{}
\expandafter\ifx\csname url\endcsname\relax
  \def\url#1{\texttt{#1}}\fi
\expandafter\ifx\csname urlprefix\endcsname\relax\def\urlprefix{URL }\fi
\expandafter\ifx\csname href\endcsname\relax
  \def\href#1#2{#2} \def\path#1{#1}\fi

\end{thebibliography}


\begin{thebibliography}{10}
\expandafter\ifx\csname url\endcsname\relax
  \def\url#1{\texttt{#1}}\fi
\expandafter\ifx\csname urlprefix\endcsname\relax\def\urlprefix{URL }\fi
\expandafter\ifx\csname href\endcsname\relax
  \def\href#1#2{#2} \def\path#1{#1}\fi

\bibitem{chen_apb_2010}
J.~Chen, A.~Hangauer, R.~Strzoda, M.-C. Amann,
  \href{http://dx.doi.org/10.1007/s00340-010-4011-0}{{VCSEL}-based
  calibration-free carbon monoxide sensor at 2.3 $\mu$m with in-line reference
  cell}, Appl. Phys. B (2010) 1--910.1007/s00340-010-4011-0.
\newblock \href {http://dx.doi.org/10.1007/s00340-010-4011-0}
  {\path{doi:10.1007/s00340-010-4011-0}}.
\newline\urlprefix\url{http://dx.doi.org/10.1007/s00340-010-4011-0}

\bibitem{boehm_jcg_2010}
G.~Boehm, A.~Bachmann, J.~Rosskopf, M.~Ortsiefer, J.~Chen, A.~Hangauer,
  R.~Meyer, R.~Strzoda, M.-C. Amann,
  \href{http://www.sciencedirect.com/science/article/pii/S0022024810011735}{Comparison of {InP-} and {GaSb-based} {VCSEL}s emitting at 2.3 $\mu$m suitable
  for carbon monoxide detection}, Journal of Crystal Growth 323~(1) (2011) 442
  -- 445, proceedings of the 16th International Conference on Molecular Beam
  Epitaxy (ICMBE).
\newblock \href {http://dx.doi.org/DOI: 10.1016/j.jcrysgro.2010.11.174}
  {\path{doi:DOI: 10.1016/j.jcrysgro.2010.11.174}}.
\newline\urlprefix\url{http://www.sciencedirect.com/science/article/pii/S0022024810011735}

\bibitem{cerutti_jcg_2009}
L.~Cerutti, A.~Ducanchez, G.~Narcy, P.~Grech, G.~Boissier, A.~Garnache,
  E.~Tourni{\'e}, F.~Genty, {G}a{S}b-based {VCSEL}s emitting in the
  mid-infrared wavelength range (2-3 $\mu$m) grown by {MBE}, J. Cryst. Growth
  311~(7) (2009) 1912--1916.
\newblock \href {http://dx.doi.org/DOI: 10.1016/j.jcrysgro.2008.11.026}
  {\path{doi:DOI: 10.1016/j.jcrysgro.2008.11.026}}.

\bibitem{ducanchez_el_2009_2}
A.~Ducanchez, L.~Cerutti, P.~Grech, F.~Genty, E.~Tourni{\'e}, Mid-infrared
  {GaSb}-based {EP-VCSEL} emitting at 2.63 $\mu$m., Electron. Lett. 45~(5)
  (2009) 265--267.

\bibitem{bachmann_njp_2009}
A.~Bachmann, S.~Arafin, K.~Kashani-Shirazi,
  \href{http://stacks.iop.org/1367-2630/11/i=12/a=125014}{Single-mode
  electrically pumped {GaSb}-based {VCSEL}s emitting continuous-wave at 2.4 and
  2.6 $\mu$m}, New J. Phys. 11~(12) (2009) 125014.
\newline\urlprefix\url{http://stacks.iop.org/1367-2630/11/i=12/a=125014}

\bibitem{huang_np_2007}
M.~Huang, Y.~Zhou, C.~Chang-Hasnain,
  \href{http://dx.doi.org/10.1038/nphoton.2006.80}{A surface-emitting laser
  incorporating a high-index-contrast subwavelength grating}, Nat. Photon.
  1~(2) (2007) 119--122.
\newblock \href {http://dx.doi.org/10.1038/nphoton.2006.80}
  {\path{doi:10.1038/nphoton.2006.80}}.
\newline\urlprefix\url{http://dx.doi.org/10.1038/nphoton.2006.80}

\bibitem{mateus_ptl_2004}
C.~Mateus, M.~Huang, Y.~Deng, A.~Neureuther, C.~Chang-Hasnain, Ultrabroadband
  mirror using low-index cladded subwavelength grating, IEEE Photon. Technol.
  Lett. 16~(2) (2004) 518--520.
\newblock \href {http://dx.doi.org/10.1109/LPT.2003.821258}
  {\path{doi:10.1109/LPT.2003.821258}}.

\bibitem{chevallier_apa_2010}
C.~Chevallier, N.~Fressengeas, F.~Genty, J.~Jacquet, Optimized sub-wavelength
  grating mirror design for mid-infrared wavelength range, Appl. Phys. A-
  Mater. 103~(4) (2011) 1139--1144.
\newblock \href {http://dx.doi.org/10.1007/s00339-010-6059-4}
  {\path{doi:10.1007/s00339-010-6059-4}}.

\bibitem{karagodsky_oe_2010}
V.~Karagodsky, F.~G. Sedgwick, C.~J. Chang-Hasnain, Theoretical analysis of
  subwavelength high contrast grating reflectors, Opt. Express 18~(16) (2010)
  16973--16988.
\newblock \href {http://dx.doi.org/10.1364/OE.18.016973}
  {\path{doi:10.1364/OE.18.016973}}.

\bibitem{wang_ol_2010}
J.~Wang, Y.~Jin, J.~Shao, Z.~Fan,
  \href{http://dx.doi.org/10.1364/OL.35.000187}{Optimization design of an
  ultrabroadband, high-efficiency, all-dielectric grating}, Opt. Lett. 35~(2)
  (2010) 187--189.
\newblock \href {http://dx.doi.org/10.1364/OL.35.000187}
  {\path{doi:10.1364/OL.35.000187}}.
\newline\urlprefix\url{http://dx.doi.org/10.1364/OL.35.000187}

\bibitem{shokooh-saremi_oe_2008}
M.~Shokooh-Saremi, R.~Magnusson,
  \href{http://www.opticsexpress.org/abstract.cfm?URI=oe-16-22-18249}{Wideband
  leaky-mode resonance reflectors: Influence of grating profile and sublayers},
  Opt. Express 16~(22) (2008) 18249--18263.
\newblock \href {http://dx.doi.org/10.1364/OE.16.018249}
  {\path{doi:10.1364/OE.16.018249}}.
\newline\urlprefix\url{http://www.opticsexpress.org/abstract.cfm?URI=oe-16-22-18249}

\bibitem{soref_joa_2006}
R.~A. Soref, S.~J. Emelett, W.~R. Buchwald,
  \href{http://stacks.iop.org/1464-4258/8/i=10/a=004}{Silicon waveguided
  components for the long-wave infrared region}, Journal of Optics A: Pure and
  Applied Optics 8~(10) (2006) 840.
\newline\urlprefix\url{http://stacks.iop.org/1464-4258/8/i=10/a=004}

\bibitem{mrcwa}
H.~Rathgen, mrcwa 20080820, http://mrcwa.sourceforge.net/ (February 2010).

\bibitem{moharam_josaa_1995}
M.~G. Moharam, D.~A. Pommet, E.~B. Grann, T.~K. Gaylord,
  \href{http://josaa.osa.org/abstract.cfm?URI=josaa-12-5-1077}{Stable
  implementation of the rigorous coupled-wave analysis for surface-relief
  gratings: enhanced transmittance matrix approach}, J. Opt. Soc. Am. A 12~(5)
  (1995) 1077--1086.
\newline\urlprefix\url{http://josaa.osa.org/abstract.cfm?URI=josaa-12-5-1077}

\bibitem{openopt}
D.~L. Kroshko, Open{O}pt 0.27, http://openopt.org/ (December 2009).

\end{thebibliography}

\listoffigures
\listoftables

\end{document}